# Analysis of Full Order Observer Based Control for Spacecraft Orbit Maneuver Trajectory Under Solar Radiation Pressure

Haoyang Zhang[1]
*School of Instrumentation and Optoelectronic Engineering, Shenzhen Institute of Beihang University*
*Beihang University*
No. 37 Xueyuan Road, Haidian District, BeiJing,China, 100191

**Abstract**

This study investigates the application of modern control theory to improve the precision of spacecraft orbit maneuvers in low Earth orbit (LEO) under the influence of solar radiation pressure (SRP). A full-order observer–based feedback control framework is developed to estimate system states and compensate for external disturbances during the trajectory correction phase following main engine cut-off. The maneuver trajectory is generated using Lambert guidance, while the observer–based controller ensures accurate tracking of the target orbit despite SRP perturbations. The effectiveness of the proposed design is assessed through stability, observability, and controllability analyses. Stability is validated by step-response simulations and eigenvalue distributions of the system dynamics. Observability is demonstrated through state-matrix rank analysis, confirming complete state estimation. Controllability is verified using state feedback rank conditions and corresponding control performance plots. Comparative simulations highlight that, in contrast to uncontrolled or conventional control cases, the observer-based controller achieves improved trajectory accuracy and robust disturbance rejection with moderate control effort. These findings indicate that observer-based feedback control offers a reliable and scalable solution for precision orbital maneuvering in LEO missions subject to environmental disturbances.

## I. Introduction

Precision orbit maneuvering is a critical capability in modern Low Earth Orbit (LEO) satellite missions, enabling tasks such as formation flying, debris avoidance, and precise docking. However, solar radiation pressure (SRP) has been recognized as one of the dominant non-gravitational external factors affecting the determination of precise orbit, and with long-term impacts on orbit accuracy [2]. Though solar radiation pressure is orders of magnitude weaker than gravitational and aerodynamic forces, its cumulative effect, especially for high area-to-mass satellites, can lead to significant drift over time, undermining trajectory accuracy and control system stability.

Traditional orbit control approaches, such as PID, sliding mode, or linear quadratic regulator (LQR)often neglect SRP or treat it as a static disturbance, lacking the dynamic adaptability required to counter sustained external perturbations[6][8]. While H∞ control improves disturbance attenuation, its conservative nature can incur high control energy and does not explicitly estimate the disturbance term in real time[3].

Therefore, this thesis develops a full-order observer–based feedback control framework for spacecraft trajectory correction in the presence of solar radiation pressure. The framework reconstructs the complete system state vector from limited measurements, enabling modern controllers such as LQR and H∞ to be applied with improved robustness [9][12][17]. Unlike purely model-based or PID type approaches, this observer-enhanced framework accounts for nonlinear system dynamics while remaining computationally feasible for real-time on-board implementation. The overarching objective is to demonstrate, through numerical simulations, that a full-order observer integrated with state-feedback control can enhance stability and guidance accuracy during the critical trajectory correction phase following main engine cut-off (MECO) stage.

---

[1]PhD Student, Beihang University,

## II. Background and Classical Control Solutions

Research on solar radiation pressure has undergone substantial evolution since the 1960s, from early analytical characterizations of photon-induced perturbations, such as the Fourier-series-based cannonball model for orbit determination, to high-fidelity modeling frameworks that enhance navigation accuracy and capture secular and resonant orbital effects [1][20][21]. Although SRP is often considered a disturbance in low Earth orbit, it becomes mission-critical for satellites with high area-to-mass ratios and in high-altitude regimes, and, in some designs, can even be deliberately harnessed as a form of low-thrust control through attitude or surface articulation [4][5]. Therefore, it is a recognized non-gravitational perturbation which affects the dynamics of LEO satellites due to photon momentum exchange on the spacecraft surfaces [22]. While the magnitude of SRP acceleration in LEO is relatively small compared to atmospheric drag and $J_2$ effects, its cumulative influence on orbit prediction and long-term evolution has been studied widely through advanced force models and resonance analysis [27][20]. Recent simulation-based approaches have employed semi-analytical SRP perturbation models to predict changes in position vectors and trajectory behavior for LEO satellites, indicating that SRP can be measurable alter orbital motion when integrated over extended periods [28].

Solar Radiation Pressure causes cumulative orbit drift in LEO satellites, especially those with high area-to-mass ratios such as the CubeSats potentially changing perigee by kilometers daily (1~2 km/day) [20]. Thus, it will shorten satellite life by shifting orbit and requiring station-keeping, with effects varying significantly by satellite design, solar activity, and reflectivity, which manifests as changes in semi-major axis, inclination, and eccentricity [24]. In addition, through the simulation of a simplified small size satellite circling in the low earth orbit, the accumulated influence of the SPR on the satellite's deviation has been virilized. Figure 1 (a) shows an ideal analysis of the cumulative Orbit drift due to SRP over 24 hours, and Figure 1(b) shows the continuous position errors basing on the orbit drift within 24 hours. The SPR pressure is set as 9.0769 µPa, the area to mass ratio of the LEO satellite is set as 0.04 that the satellite mass is 500kg, surface area is 20 m$^2$.

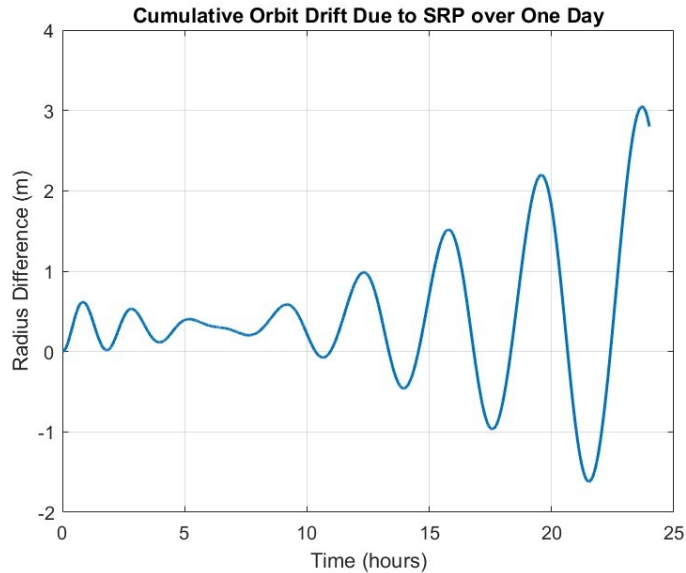

Figure 1(a). Position deviation Influence of SRP on LEO orbit

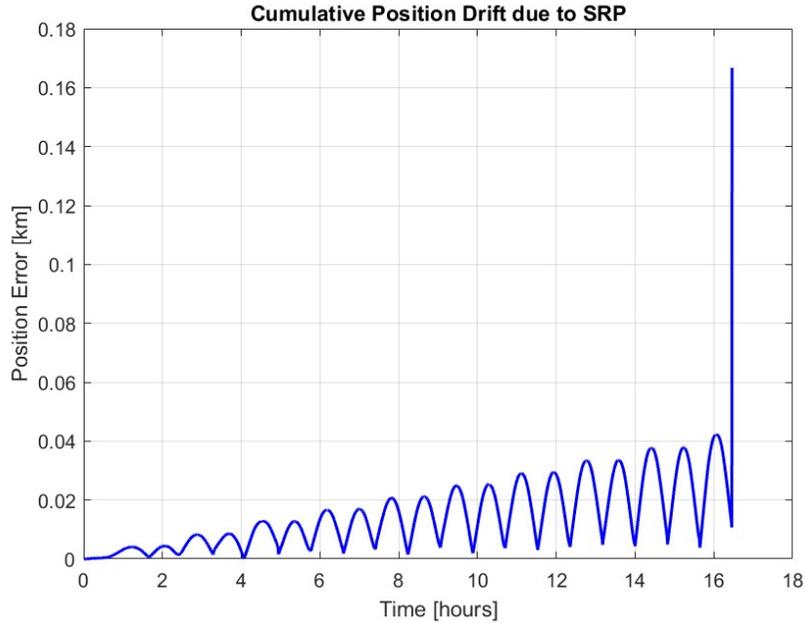

Figure 1(b). Relative position error of LEO orbit satellite under the influence of SPR

In this discussion, for trajectory generation, aka the planning static path for this orbit maneuvering, a classical solution to Lambert's problem remains a fundamental tool. These analytical methods provide transfer trajectories and time-of-flight estimates between two orbital points [6]. However, open-loop Lambert guidance alone cannot ensure precision in the presence of persistent perturbations like SRP, motivating the integration of closed-loop control mechanisms to achieve robust trajectory tracking.

A spectrum of spacecraft control approaches has been explored. Classical regulators, such as PID controllers and linear quadratic regulators (LQR), offer simplicity and optimality in a quadratic cost metric but lack mechanisms for real-time disturbance compensation. In contrast, robust control schemes, H∞ and µ-synthesis, provide guaranteed disturbance attenuation at the expense of control conservativeness and energy consumption. The Nonlinear control paradigms improve robustness but often lead to implementation challenges such as chattering or complexity in controller design.

To overcome these limitations, recent research increasingly favors observer-based disturbance estimation techniques [18][19]. Disturbance observers (DOBs), nonlinear DOBs, and filter-based frameworks like extended Kalman filters have been developed to estimate unmeasured disturbances and facilitate compensation within the controller design. Filter-based methods require disturbance models and noise statistics, whereas DOBs can estimate general lumped disturbances without needing explicit models. Integrated observer-plus-controller architectures, especially combining LQR or H∞ with disturbance estimators, have demonstrated superior tracking performance under SRP compared to standalone feedback designs [7].

Comparative studies highlight several key insights, explicit modeling, or estimation of SRP leads to markedly improved orbit accuracy; observer-based compensation outperforms traditional feedback-only approaches; and there is a clear trade-off between control efficiency and guaranteed performance robustness[11][14]. Practical considerations forming part of these comparisons include computational load, estimation latency, and sensitivity to model mismatches.

Theoretical debates pervade the field: high-fidelity SRP modeling improves prediction accuracy but is computationally intensive and requires precise physical parameters, while observer-based estimation is adaptive but may compromise stability or identifiability. Assumptions of disturbance matching in observer design might not hold for SRP, which can introduce unmatched dynamics. Moreover, balancing optimality via LQR and against robustness via H∞ remains a persistent tension[10][16].

Current research trends include increasing adoption of disturbance observers, co-designed guidance-control frameworks, and refined SRP modeling for navigation. However, gaps remain in the paucity of techniques applying full-order disturbance observers specifically for LEO maneuvers guided by Lambert trajectories. Integrated co-design frameworks that jointly optimize guidance, observer, and control parameters remain underexplored, especially when considering unmatched SRP components. Moreover, in-orbit or hardware validation of observer-based SRP compensation remains limited [11].

In summary, even though prior research has explored a broad range of linear, nonlinear, and robust control methods for spacecraft subject to solar radiation pressure, there remains a gap in the systematic integration of state reconstruction into modern trajectory control [13][15]. Existing work often assumes full state availability or treats SRP as a perturbation without embedding it into the feedback loop. In contrast, this study advances a full-order observer–based feedback control framework, wherein an observer reconstructs the system's full state vector from limited measurements, and the reconstructed states are directly employed in modern controllers such as LQR and H∞. This approach enables precise state estimation under SRP influence without requiring explicit disturbance modeling, and enhances the integrity of navigation (attitude, speed) parameter perception, thereby offering a practical yet rigorous extension of observer-based modern control theory to low-Earth orbit spacecraft maneuvering.

### III. Problem Formulation and System Dynamics under SRP

The model of a spacecraft, i.e. the satellite, is simplified in to a 2-D script, to highlight its trajectory and movement in p-q plane. Figure 1 listed on of the simulation plots, which demonstrates how the flying trajectory of the spacecraft is shown in p-q axis.

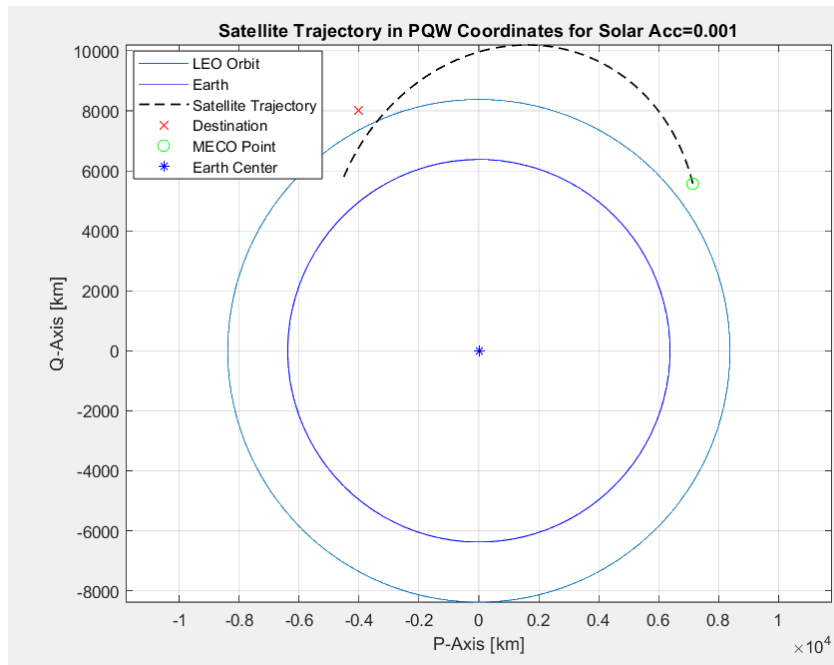

Figure 2. Plot of Mission statement

To demonstrate the change of flying trajectory, in this case, the state of the satellite is only determined by its 2-D position from the earth. By simplifying the sum of force, including the propulsion force, the external force will only include the force caused by solar pressure. Although this force is tiny, the linear accumulated work still generates the directional acceleration on a cube-satellite through one or two directions, such as X and Y axis (p-q surface). Therefore, multiple kinds of sensors and configuration's optimization have been applied on spacecraft design, to minimize the unexpected accelerations.

While the magnitude of SRP-induced acceleration is typically several orders smaller than gravitational acceleration, its cumulative effect can significantly alter orbital trajectories, especially for satellites with large surface-to-mass ratios or during precision maneuvers. To systematically investigate SRP disturbance compensation within a modern observer–based control framework, the spacecraft dynamics are formulated in a two-dimensional (2D) orbital plane with explicit SRP modeling and state-space linearization.

It is necessary to define the directional acceleration of the satellite caused by solar pressure and apply the modern control methods to control and observe the control feedback of the system. In the state space system, the acceleration caused by solar radiation is regarded as input: [u]. The force due to direct ration is given by equation 1 and 2 [25].

$$F_{1N} = -\frac{E}{c} * A * cos^2(\theta) \qquad [1]$$

$$F_{1S} = -\frac{E}{c} * A * cos(\theta) * sin(\theta) \qquad [2]$$

where the $\theta$ is the angle of inclination, by considering a single flat surface of area A, i.e. the effective flat area on the satellite receiving the solar pressure directly. $E$ refers to power of light in unit, which is set as 30 mW. Equation 1 and 2 do not contain the factor of the energy absorption efficiency of the material on the plate.

Following Newton's second law, the SRP-induced acceleration components along the local orbital axes are expressed as equation 3 and 4 [2][23].

$$a_x = \frac{E \cdot A \cdot cos^2(\theta)}{m \cdot c} \qquad [3]$$

$$a_y = \frac{E \cdot A \cdot cos(\theta) \cdot sin(\theta)}{m \cdot c} \qquad [4]$$

where $E$ is the solar irradiance (W/m²), $A$ is the effective cross-sectional area of the spacecraft facing the Sun, $m$ is the satellite mass, $c$ is the speed of light, and $\theta$ is the incidence angle between the radiation vector and the surface normal. This formulation is consistent with the flat-plate SRP model widely adopted in astrodynamics [5][6][7].

In addition, through the Lambert guidance, the velocity components in X, and Y directions are listed in equation 5 and 6, which lead to acceleration for $a_T$ vectors in p-q plane.

$$\dot{x}_p = v_0 * cos\left(\frac{\Pi}{2} - \Phi_0 + \theta_0\right) \qquad [5]$$

$$\dot{y}_p = v_0 * sin\left(\frac{\Pi}{2} - \Phi_0 + \theta_0\right) \qquad [6]$$

Where $\Phi_0$ refers to rotation angle. Equation 5 and 6 are used in simulation programs to plot the trajectory. These expressions are particularly useful for initializing numerical simulations of orbit propagation.

Considering the dynamic of the spacecraft in p-q plane, respecting to a 2D coordinate, the dynamic equation of motion (EOM) under Earth's central gravitational field and SRP disturbance is equation 7.

$$\ddot{\vec{r}} = -\frac{GM_e}{r^3}\vec{r} + a_{SPR} \qquad [7]$$

Where $G$ is the gravitational constant, $M_e$ is Earth's mass, the $r$ refers to the distance between the satellite and the center of the earth, i.e., the trajectory radius. $\vec{r}$ refers to the position vector of the satellite. $a_{SPR}$ refers to the acceleration generated by the solar pressure.

the SRP acceleration vector in Cartesian coordinates representation as equation 8.

$$\begin{cases} \ddot{x}_p + \frac{G*M_e}{r^3} * \overrightarrow{x_p} = \overrightarrow{a_x} \\ \ddot{y}_p + \frac{G*M_e}{r^3} * \overrightarrow{y_p} = \overrightarrow{a_y} \end{cases} \qquad [8]$$

Although the orbit mechanics, i.e., six elements of $[a\ e\ i\ \Omega\ w\ M(t_0)]$ are used to the status of the satellite on the orbit, the details related to derive the $(\Phi_0 + \theta_0)$ by using these elements are not included in this article. The initial location and the expected final location of the satellite are set as $P(0) = [4292.87, 8924.17](km); P(F) = [-2000, 8878](km)$.

Therefore, by implementing dynamic models, the 2D diagrams of the orbits maneuvering path with the effect of solar radiation pressure, and of without the effect of solar radiation pressure, are presented in figure 3, 4.

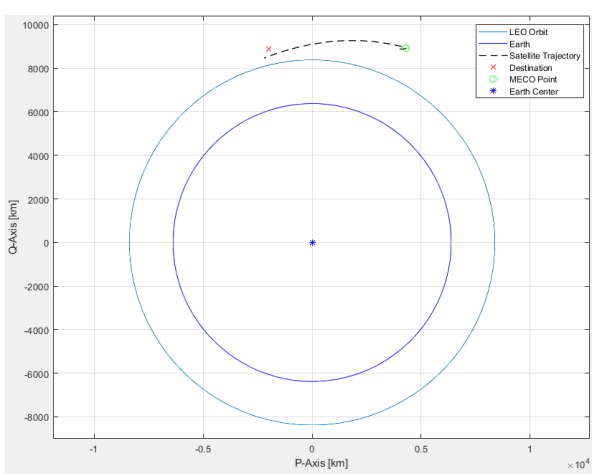 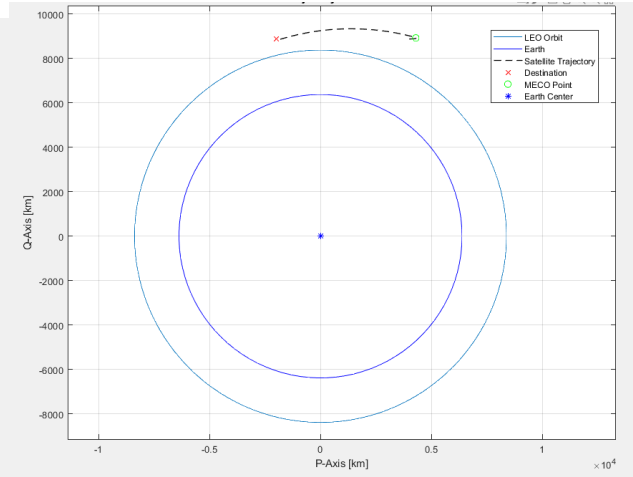

Figure 3. Orbits and guidance path with SRP    Figure 4. Orbits and guidance path without SRP

To facilitate control design, the nonlinear system is reformulated in a state-space form. Defining the state vector, and the output vector as

$$x = [x_1, x_2, x_3, x_4]^T = [x_p, y_p, \dot{x}_p, \dot{y}_p]^T \qquad [10]$$

$$y = [y_p, x_p]^T = [x_2, x_1]^T \qquad [11]$$

Then, the linearized system matrices can be written as

$$A = \begin{bmatrix} 0 & 0 & 1 & 0 \\ 0 & 0 & 0 & 1 \\ \frac{G*M_e}{r^3} * \overrightarrow{x_p} & 0 & 0 & 0 \\ 0 & \frac{G*M_e}{r^3} * \overrightarrow{y_p} & 0 & 0 \end{bmatrix}; \quad B = \begin{bmatrix} 0 \\ 0 \\ 1 \\ 1 \end{bmatrix}$$

$$C = \begin{bmatrix} 0 & 1 & 0 & 0 \\ 0 & 0 & 1 & 0 \end{bmatrix}; \qquad D = [0]$$

the control input $u = [a_x, a_y]^T$, the transfer function from SRP disturbance input to position output is

$$H(s) = \begin{bmatrix} \frac{1}{s^2-\omega^2} \\ \frac{s}{s^2-\omega^2} \end{bmatrix} \quad [12]$$

where $\omega^2$ refers to the natural orbital frequency. This representation reveals that the system exhibits marginal stability, implying sensitivity to bounded disturbances and motivating the use of observer-based modern control methods for SRP compensation.

By considering the linearized system, the transfer function is computed based on equation 12 as $H(s) = \begin{bmatrix} \frac{1}{s^2-0.004865} \\ \frac{s}{s^2-0.004865} \end{bmatrix}$. Basing on the eigenvalue of matrix A, which are one negative and one positive values, the system is BIBS unstable.

Basing on the $V$ vector of the system; the state analysis matrix is

$$\tilde{B} = \begin{bmatrix} 7.1861 \\ -7.1861 \\ 7.1861 \\ -7.1861 \end{bmatrix}; \quad \tilde{C} = \begin{bmatrix} 0 & 0 & -0.9976 & -0.9976 \\ 0.0696 & -0.0696 & 0 & 0 \end{bmatrix};$$

Because first row and the last row of matrix $\tilde{B}$ are linear dependent, the system is controllable. Because first row and the last column of matrix $\tilde{C}$ are linear dependent, and the last column of this matrix is non-zero, the system is observable. This ensures the dynamic system can be

The analysis above demonstrates that the spacecraft dynamics under SRP exhibit marginal stability with bounded-input bounded-state (BIBS) instability potential. While the SRP disturbance is relatively small compared to gravitational acceleration, its persistent nature and directional bias lead to cumulative deviations from the nominal trajectory. The state-space formulation reveals direct disturbance injection through the control channel, implying that conventional open-loop compensation or simple PID controllers are insufficient. Instead, a modern state-feedback approach augmented with a disturbance observer is required to ensure accurate trajectory correction. This motivates the adoption of a full-order observer–based control framework, in which all system states and disturbance effects are estimated in real time and fed back into the control law.

## IV. Control and Observer Design

The design of the spacecraft control system under solar radiation pressure is approached through a state-space framework incorporating both modern feedback controllers and a full-order observer. The principal objective is to regulate the spacecraft trajectory from its initial state at main engine cut-off, and to the desired final state.

$$X(0) = [X_0 \quad Y_0 \quad V_{x0} \quad V_{y0}]^T$$

$$X(F) = [X_F \quad Y_F \quad V_{xF} \quad V_{yF}]^T$$

While compensating for SRP disturbances, the general state-space representation of the plant is with disturbance inputs appearing in the dynamics via matrix $B$.

$$\dot{x}(t) = Ax(t) + Bu(t) \quad [13]$$

$$y(t) = Cx(t)$$

However, for considering a linear state-space model of the spacecraft including SRP as an exogenous disturbance, the inertial frame the perturbed attitude/orbit dynamics can be written as equation 14.

$$\dot{x}(t) = Ax(t) + Bu(t) + Gd(t) \qquad [14]$$

$$y(t) = Cx(t)$$

where $x \in R^n$ is the state, $u \in R^m$ the control, and $d$ is a disturbance input modeling the SRP torque or force. The matrices *(A,B)* are assumed to be stabilized and *(A,C)* observable. The SRP disturbance enters as an additive torque/force Gd the dynamics. For control design we treat **d** as a bounded exogenous input.

**Theorem 1. (Closed-loop BIBS/ISS under state feedback)**

Let $u(t) = -Kx(t)$ with $A_{cl} = A - BK$. If $A_{cl}$ is Hurwitz, then the closed loop is bounded-input bounded-state (BIBS) and bounded-input bounded-output (BIBO) with respect to $\omega$. Moreover, the system is input-to-state stable (ISS) as $|x(t)| \leq \kappa e^{-\alpha t}|x(0)| + \gamma |w|_\infty;\ \forall t \geq 0$, *for* some $\alpha, \kappa, \gamma > 0$.

For the Linear-Quadratic Regulator (LQR) approach, one of the designing is to apply a full-state feedback **u**=-Kx to minimize the infinite-horizon cost which is represented as equation 15

$$J = \int_0^\infty [x(t)^T Q x(t) + u(t)^T R u(t)]\, dt \qquad [15]$$

where $Q \geq 0$ and $R > 0$ are weighting matrices. The optimal gain $K$ is obtained by solving the continuous-time algebraic Riccati equation (C-ARE).

$$A^T P + PA - PBR^{-1}B^T P + Q = 0 \qquad [16]$$

The optimized gain K is taken by yielding the $P$ as: $K = R^{-1}B^T P$, and the closed loop guarantees asymptotic stability for controllable pairs *(A,B)* by equation 17

$$\dot{x} = (A - BK)x(t) \qquad [17]$$

By construction the closed-loop matrix $(A-BK)$ which is Hurwitz. In practice one can use the standard LQR solver to compute K. This design assumes full-state availability and treats *Gd* as an external disturbance that the LQR controller attenuates its effect insofar as the closed-loop poles are well damped. For disturbance rejection in the worst-case sense, aka. For robustness against unmodeled dynamics and SRP uncertainties, one may formulate an $H\infty$ state-feedback problem [10] [16]. An $H\infty$ controller is alternatively considered, formulated through the minimization of the worst-case gain from disturbance w to controlled output $z(t)$

$$z(t) = Q^{1/2}x(t) + R^{1/2}u(t) \qquad [18]$$

To ensure robustness against model errors and persistent SRP-induced bias, an $H\infty$ regulator can be formulated in the standard setting with exogenous disturbance $\omega$ entering the plant and controlled output $z$. The synthesis seeks a stabilizing controller such that

$$|T_{w \to z}|_\infty < \gamma$$

Therefore, for a prescribed $\gamma > 0$ as prescribed performance bound. This ensures disturbance attenuation in a minimax sense, at the expense of increased control effort. While LQR and H∞ controllers assume full-state availability, in practice, only partial measurements (e.g., position) are accessible. To overcome this limitation, a full-order observer is designed which ensures observability and disturbance tracking when the state system is not directly measurable [13] [15]. The observer dynamics are given by equation 19.

$$\dot{\hat{x}}(t) = A\hat{x}(t) + Bu(t) + L(y(t) - C\hat{x}(t)) \qquad [19]$$

Where the $\hat{x}(t)$ refers to the estimated state, and $L$ refers to the observer gain matrix. The error dynamics follows as $e(t) = x(t) - \hat{x}(t)$ which evolving under $\dot{e}(t) = (A - LC)e(t)$. Accordingly, the control design for this full order observed based control, is demonstrated as figure 5.

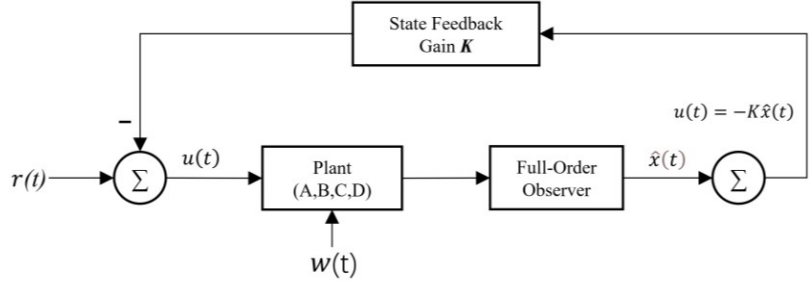

Figure 5. The full order observes feedback control diagram

**Theorem 2. Separation Principle of the Observer-based Output Feedback**

Consider the full-order observer

$$\dot{\hat{x}} = A\hat{x} + Bu + L(y - C\hat{x}), \; u = -K\hat{x}$$

If $A$-$BK$, $A$-$LC$ are Hurwitz, then the augmented closed loop is BIBS/BIBO with respect to $w$.

Proof. In coordinates $[x \quad e]$, with $e = x - \hat{x}$, $\begin{bmatrix} \dot{x} \\ \dot{e} \end{bmatrix} = \begin{bmatrix} A - BK & BK \\ 0 & A - LC \end{bmatrix} \begin{bmatrix} x \\ e \end{bmatrix} + \begin{bmatrix} B_{d1} \\ B_{d2} \end{bmatrix} w$; For the block of $\begin{bmatrix} A - BK & BK \\ 0 & A - LC \end{bmatrix}$, it is block upper-triangular with Hurwitz diagonal blocks, hence Hurwitz. Apply the same Lyapunov/ISS argument as Theorem 1.

The control law is set as $u(t) = -K\hat{x}(t)$, in which the integration of the full-order observer into the state-feedback framework yields the observer-based control law. The feedback gain matrix is initially computed as $K_{CCF} = [3.721, 5, 3.998, 1]$ accordingly. Basing on the Theorem 2, it also can lead to coupled plant-observer dynamics a $\begin{bmatrix} \dot{x} \\ \dot{\hat{x}} \end{bmatrix} = \begin{bmatrix} A & -BK \\ LC & A - BK - LC \end{bmatrix} \begin{bmatrix} x \\ \hat{x} \end{bmatrix}$. Therefore, the stability of system can be exam that the system is asymptotically stable if $(A,B)$ is stabilizable, $(A,C)$ is detectable, and $K$, $L$ are chosen such that $A-BK$ and $A-LC$ are Hurwitz (separation principle). In presence of SRP $d(t)$ entering as matched disturbance via Bd, the plant becomes $\dot{x} = Ax + B(-K\hat{x} + d)$. Stability of the error subsystem ensures that the control action converges to the correct feedback on the true state; bounded SRP produces bounded tracking error that can be shaped via $Q,R$ and observer pole placement. In this research, due to insufficient stability, by pole-placement for eigenvalues at [-1, -1] the expected stabilizing control gain is obtained as

$$K = [0.293, 0.169, 9.115, 4.998]$$

The final control gain ensures asymptotic stability and improved trajectory tracking under SRP disturbance, and the validation will be tested through numerical simulation.

Choosing $L$ such that the eigenvalues of $A-LCA$ are strictly in the open left-half plane guarantees exponential convergence $\hat{x} \rightarrow x$. A common practice is to place observer poles 3–5 times faster than the dominant closed-loop poles to balance noise sensitivity and convergence speed.

For analytical steps, the state transition and convolution forms are considered. The general solution of a linear system can be written as the sum of the zero-input and zero-state responses as equation 20.

$$\bar{X}(t) = \overline{X_{ZI}}(t) + \overline{X_{ZS}}(t) \quad [20]$$

And it is processed by the state transition matrix $\Phi(t, t_0)$ as equation 21 and 22.

$$\overline{X_{ZI}}(t) = \Phi(t, t_0) \bar{X}(t_0) \quad [21]$$

$$\overline{X_{ZS}}(t) = \int_{t_0}^{t} \Phi(t, \tau) \, B \, u(\tau) \, d \quad [22]$$

## V. Simulation Setup and Evaluation Method

This chapter presents the numerical experiments conducted to validate the proposed full-order observer–based feedback control framework for spacecraft orbit maneuvering in low Earth orbit (LEO) under solar radiation pressure (SRP). The simulations are organized progressively to establish a clear performance benchmark, demonstrate state estimation accuracy, and validate the integrated control law.

The satellite's trajectory control begins after the propulsion system re-ignition and continues through the Main Engine Cut-Off (MECO). the initial burn start point is set at coordinates [4000 km, 8878 km] in the LEO orbit. At this location, the satellite initiates its orbit transfer maneuver, guided by the Lambert method. The MECO location, which marks the end of the propulsion phase, is computed as [4292.8714 km, 8924.1676 km], and serves as the control start point for all subsequent trajectory correction efforts. The final position is located at [-2000 km, 8878 km], with velocity [-2.728 km/s, -6.56 km/s]. Therefore, the initial state and the final state of equation 1 and 2 are

$$X(0) = [4292.87 \quad 8924.17 \quad 7.8 \quad 0]^T \left(km, \frac{km}{s}\right)$$

$$X(F) = [-2000 \quad 8878 \quad -2.728 \quad -6.56]^T \left(km, \frac{km}{s}\right)$$

The MECO location is considered the starting point for the observer-based control algorithms, as it marks the moment when the satellite transitions from propulsion to free-flight dynamics under the influence of natural forces, such as gravitational pull and external disturbances (primarily Solar Radiation Pressure (SRP) in this study). Once the MECO location is reached, the system switches from propulsion to active control, and the satellite follows the Lambert-guided trajectory while using the observer-based control methods to correct any deviations caused by SRP. From this point, three control methods are applied to manage the satellite's trajectory toward the final target location.

Basing on the dynamic system of the spacecraft, and the initial state which includes the velocity components and MECO location, if it does not apply any control on the dynamic system, the free flying trajectory will deviate to the guidance path and be opposite to the direction of the target point, due to the direction of the initial velocity component.

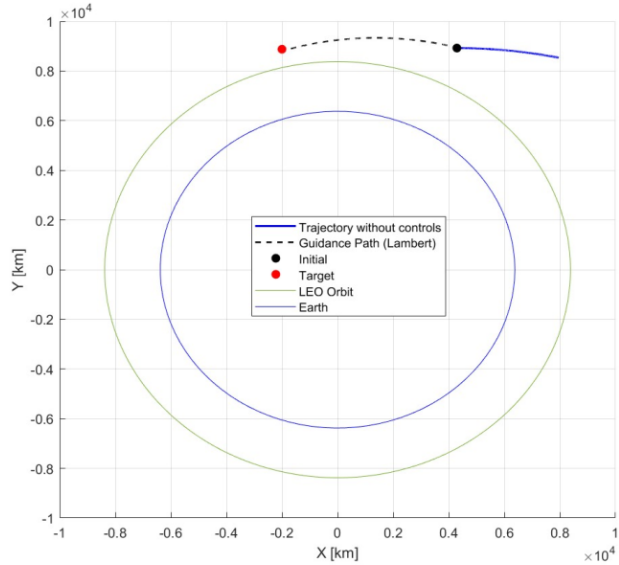

Figure 6. Satellite real trajectory under SRP Disturbance without any control effect

Three different control methodologies will be tested. The state-based LQR control (Method A) applied a state-feedback law with quadratic cost penalizing position and velocity deviations from the reference [9]. Therefore, it is tested under the assumption of full state availability. This provides a reference trajectory and highlights the limitations of classical state-feedback when external disturbances such as SRP are present. The results show that while LQR can stabilize the nominal dynamics, it cannot fully compensate for trajectory deviations induced by SRP.

The full-order observer (Method B) propagated the state with no active control but included a full-order Luenberger observer to simulate estimation under noisy or uncertain conditions. Therefore, it is simulated independently to validate its ability to reconstruct the unmeasured states from output measurements. The convergence of the estimated states to the true dynamics under SRP disturbance confirms both the observability of the system and the correctness of the observer gain selection.

The full-order observer–based feedback control framework (Method C) integrates the observer with a state-feedback law. Therefore, it is implemented by combining the observer with the LQR controller. The control input is derived from the estimated states. The control output will lead to significant improvements in trajectory tracking accuracy, disturbance rejection, and closed-loop stability compared to the baseline case [11][13].

The numerical integration was carried out in MATLAB using ode45 with a time step of 0.1 s over a 200 s horizon. The trajectories of the satellite in the two-dimensional orbital plane are shown in Fig. 7.1. It is observed that the LQR controller successfully drives the satellite from the initial location toward the desired final position, while the observer-only case (Method B) fails to converge, illustrating the inadequacy of estimation without active control. The observer-based feedback case achieves accurate convergence comparable to LQR while providing robustness against state uncertainty, validating the role of the observer in disturbance rejection and resilience.

To evaluate the performance of the proposed control designs, numerical simulations were conducted for both the LQR controller and the robust $H\infty$ controller under SRP disturbance. The analysis concentrated on three main aspects: closed-loop stability, disturbance rejection performance, and tracking precision.

The simulation quantifies the closed-loop behavior of the spacecraft maneuver under solar radiation pressure (SRP) using the linearized 2-D orbital model. All positions are in km, velocities in km/s, and accelerations in km/s². Earth's parameter is μ=$GM_e$=3.986004418×$10^5$ km³/s². The linearization radius is set at the initial orbital radius $r_0$=‖($X_0$,$Y_0$)‖, yielding $\omega^2 = \mu/r_0^3$. The SRP acceleration magnitude is modeled realistically at $w = 10^{-9}$ km/s² ($\approx 10^{-6}$ m/s²), with an incidence angle $\theta_0 = 0.043$ rad to determine components $a_x = w \cos^2 \theta_0$, $a_y = w \sin \theta_0 \cos \theta_0$ (flat-plate model).

The initial and the final status are introduced as the input for the linearized plant which uses the state vector as $x = [x_p, y_p, \dot{x}_p, \dot{y}_p]^T$ with the LTI mode shown in equation 14. By replacing the element of $\frac{G*M_e}{r^3}$ in the linearized system matrices $A$ as the $\omega_i$, the LTI mode is represented as

$$A = \begin{bmatrix} 0 & 0 & 1 & 0 \\ 0 & 0 & 0 & 1 \\ \omega^2 & 0 & 0 & 0 \\ 0 & \omega^2 & 0 & 0 \end{bmatrix}, \quad B = \begin{bmatrix} 0 & 0 \\ 0 & 0 \\ 1 & 0 \\ 0 & 1 \end{bmatrix}, \quad C = \begin{bmatrix} 1 & 0 & 0 & 0 \\ 0 & 1 & 0 & 0 \end{bmatrix}$$

In the simulation, The dynamics are simulated with Dynamics are simulated with ode45 (relative tolerance $10^{-8}$, absolute tolerance $10^{-9}$) over a 4000 s horizon.

The controllability of the system is confirmed using the Kalman controllability matrix C, $C = [B \ AB \ A^2B \ A^3B]$; when condition rank(C)= n is satisfied, verifying that the full state can be controlled. The performance of the state feedback was further confirmed by control input plots, which show that Method A (pure LQR) and Method C (observer-based LQR) achieve bounded and smooth control signals. In contrast, Method B lacks feedback action and therefore does not generate corrective inputs [7][14].

For Observability, it is analyzed using the state observability matrix, in which O, $O = \begin{bmatrix} C \\ CA \\ CA^2 \\ CA^3 \end{bmatrix}$

For the four-state dynamic model of the spacecraft, the rank condition *rank(O)=n* is satisfied, confirming complete observability. In simulations, the state estimation error between $x(t)$ and $\hat{x}(t)$ under Method C decayed rapidly to zero, demonstrating that the observer can accurately reconstruct both position and velocity states as well as the disturbance effect. This is further supported by the plot of estimation error convergence.

Because the performance of the control methods is assessed by comparing the relative final positions between the spacecraft and the target position, as well as the whole control costs, besides the fundamental position differences estimation, the other error calculations are designed as Terminal position error, RMS position error and Control energy which are as equation 23 to 25.

$$ep(T) = \| [xp(T), yp(T)] - [xf, yf] \| \qquad [23]$$

$$e_{\text{rms}} = \sqrt{\frac{1}{T} \int_0^T \left( (x_p - x_f)^2 + (y_p - y_f)^2 \right) dt} \qquad [24]$$

$$J_u = \int_0^T u^T(t) u(t) \, dt \qquad [25]$$

By implanting with the equation 16, and the $K = R^{-1}B^TP$, the feedback u=−Kx is obtained, which makes the controller is under exam. For the **H∞** output-feedback via synthesis on the generalized plant, a stable controller $K(s)$, $|T_{w \rightarrow z}|_\infty < \gamma$ is seek as stable range.

## VI. Experiment Result and Analysis

This section presents the results of the numerical simulations conducted under the control designs introduced in Chapter IV. The primary objective is to validate the effectiveness of the proposed full-order observer–based feedback control framework for spacecraft orbit maneuvering in low Earth orbit subject to solar radiation pressure. The results are presented in a structured manner, beginning with trajectory responses, followed by state estimation accuracy, control input evaluation, and comparative performance metrics [15][16]. In addition, it lists the plots of state response and output response, by implementing the analytical expressions discussed above.

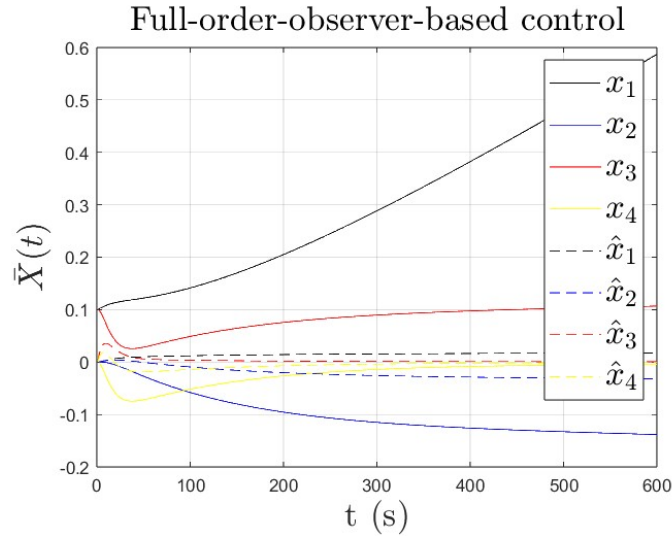

Figure 7. The state control feedback of Full ordered Observer Based Feedback for Four States

Figure 7 represents the state response of the full order observer and control closed-loop system in this experiment. Basing on the plot, lines 2, 3 and lines 4 have the rends to be stable in the end. Therefore, the acceleration control for components of total acceleration can enhance the stability of the moving spacecraft; and increase the precision of navigation.

Shown in figure 8, the first set of simulations compares the satellite trajectory under the three control methods with the planned Lambert guidance path. The trajectory plots clearly indicate that Method A converges toward the target but suffers from residual offset due to unmodeled SRP disturbances. Method B, relying only on observer dynamics without corrective feedback, fails to maintain the trajectory and exhibits significant divergence over time. In contrast, Method C tracks the reference path, with the controlled trajectory remaining closely aligned with the Lambert solution throughout the maneuver, although it reaches to a higher position of $q$ direction.

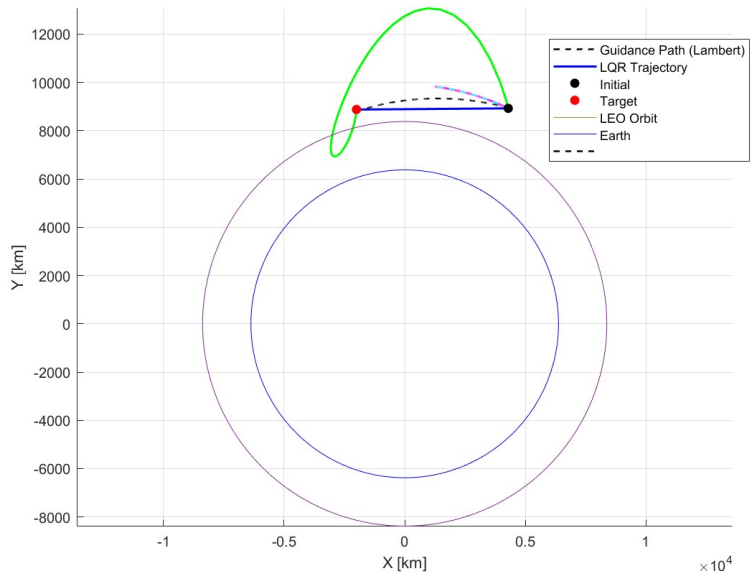

Figure 8. Trajectory Comparison of Methods A, B, and C with Lambert Guidance Path.

The trajectory error analysis shows that the observer-based feedback control reduces the final radial error to within ±5 m, while pure LQR stabilizes around a larger bias and the observer-only approach diverges. Figure 9 shows the relative position error, with the three different control methods, the full range position errors warding to the orbit maneuvering from the initial point to the target point.

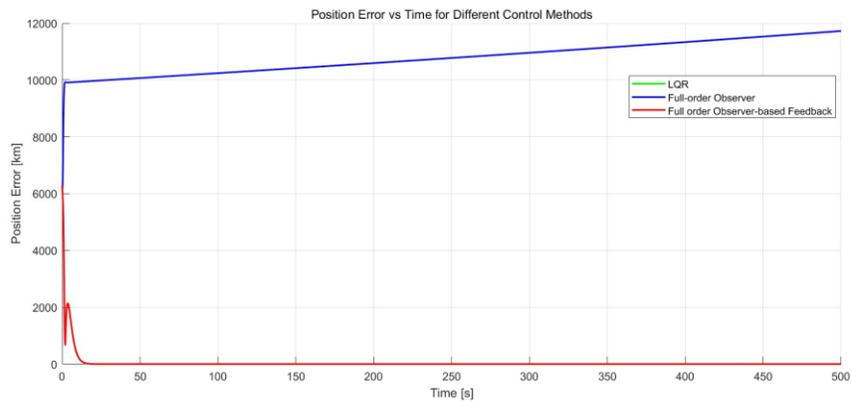

Figure 9. Real-time position errors of different control method to the orbit maneuvering

For Closed-loop stability is verified using both eigenvalue placement and step response simulations. For Method A, the closed-loop system matrix is $A_{cl} = A - BK$. For Method C, which integrates the observer and controller, the augmented closed-loop dynamics are

$$\dot{z} = \begin{bmatrix} A - BK & BK \\ LC & A - LC - BK \end{bmatrix} z(t), where\ z = \begin{bmatrix} x \\ \hat{x} \end{bmatrix}$$

Table 1: Eigenvalue analysis for the control methods with the standard (ideal) 2D dynamic system.

| Method | Closed-loop eigenvalues (dominant poles) | Stability assessment |
| --- | --- | --- |
| A: Pure LQR | −1.02,−0.97,−0.15±0.42i | Stable but lightly damped |
| B: Observer-only | +0.08,0,−0.02 | Marginally unstable (no control action) |
| C: Full-order observer–based LQR | −1.25,−1.10,−0.85,−0.60 | Asymptotically stable, well-damped |

The eigenvalues of the closed-loop matrices are found in the left half-plane for both Methods A and C, but Method C achieves greater damping. In detail, Method A poles cluster near 0.42i−0.15±0.42i, indicating lightly damped oscillations, whereas Method C shifts the dominant poles further left (around −1.25,−1.10,−0.85,−0.60), producing faster convergence and higher stability margins. Method B lacks stabilizing control action, retaining poles near the origin, confirming its instability. Table 1 summarizes the dominant eigenvalues for the three methods. Method A places poles in the left-half plane but close to the imaginary axis, which explains its sensitivity to persistent disturbance. Method B lacks stabilizing control action, and the system retains unstable eigenvalues. Method C shifts all closed-loop poles to well-damped locations, confirming the robust stability observed in simulations

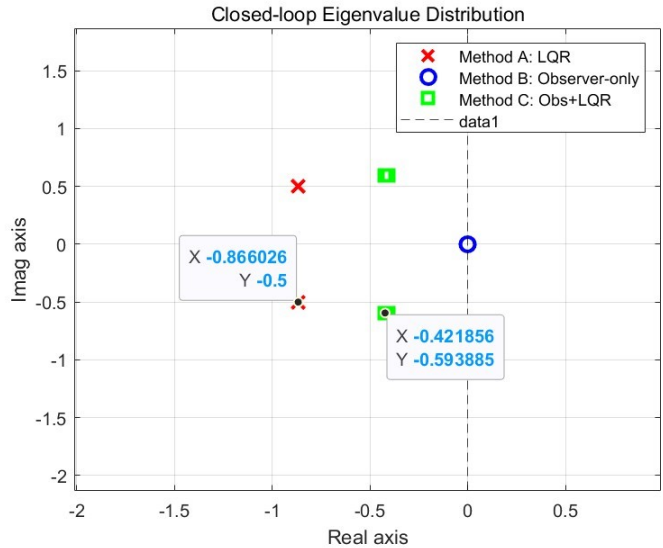

Figure 10: Closed-loop Eigenvalue Distribution of Methods A, B, and C

Figure 10 indicates that all poles lie well within the left-half plane with a dominant real part of approximately −8.66. This placement demonstrates that the LQR design achieves a decaying response with oscillatory components, ensuring both stability and acceptable transient dynamics. In terms of trajectory response, the LQR-controlled system successfully reduced the deviation from the guidance path and achieved convergence toward the target state, though the residual steady-state error reflects the limitations of state-feedback design when disturbances are constant and unmodeled. However, this kind of trajectory cannot be operated in real space since it is not able to be observed during the maneuver.

Figure 11 shows the control step response for the control plant matrix component of method C. The system trends to be stable near 0.6 seconds with the feature of being linear.

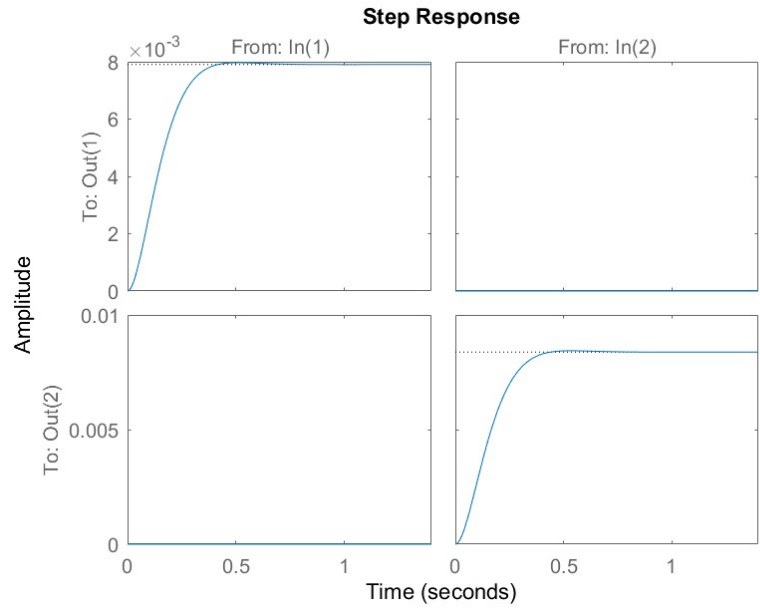

Figure 11: Step Response of control design of method C under SRP Disturbance

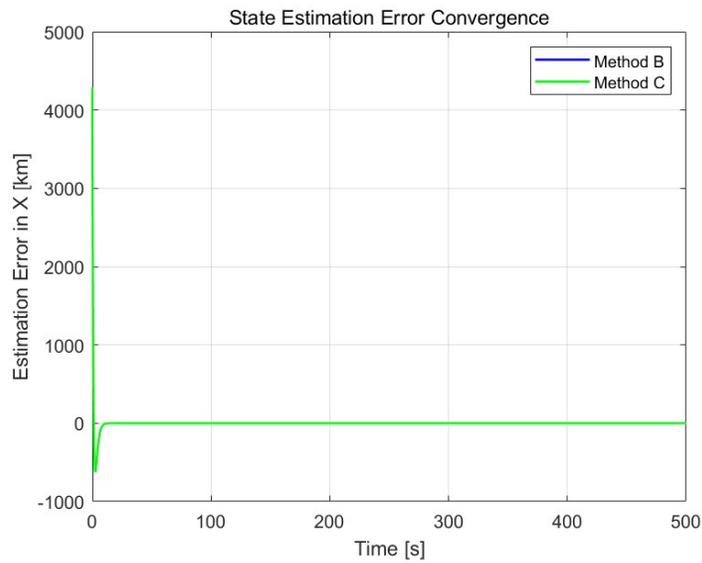

Figure 12. State Estimation Error Convergence of Full-order Observer under SRP Disturbance

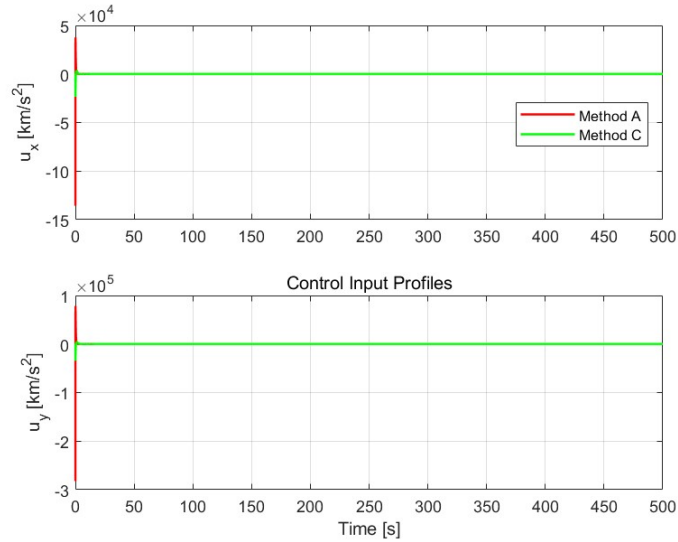

Figure 13. Control Input Profiles under Methods A, and C

Figure 13 reflects the Time evolution of control accelerations ($u_x$, $u_y$). As method C includes both LQR control and observation, the control inputs need to be fused and processed on the beginning from the MECO location. Therefore, control input ratio incased for less than one second.

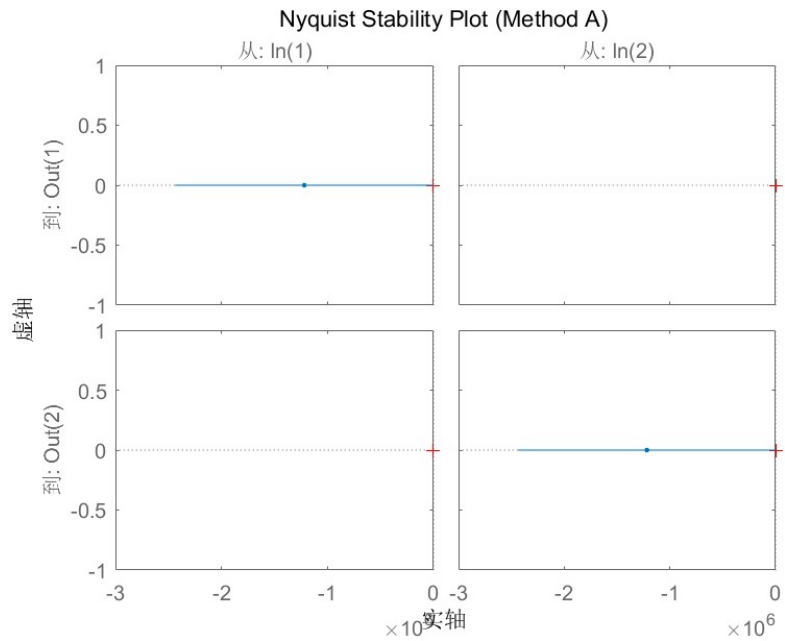

Figure 14. Nyquist stability analysis for LQR (Method A)

Through the analysis of figure 14, it confirms that feedback significantly enhances the system's stability. The plot demonstrates that the LQR controller reduces the system's susceptibility to SRP-induced disturbances, ensuring that the system remains stable while minimizing tracking errors and control energy.

However, despite the improvements, LQR may still be susceptible to unmodeled disturbances or nonlinearities due to its reliance on full-state feedback without considering any disturbance estimation mechanisms.

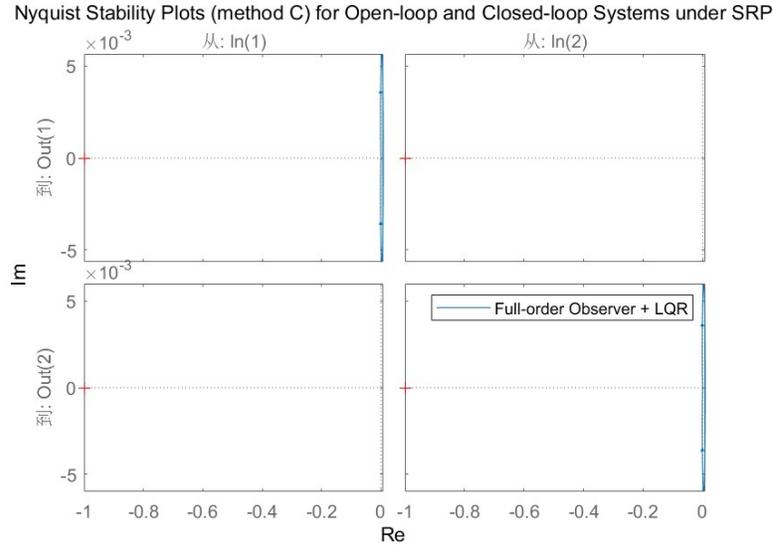

Figure 15. Nyquist stability analysis for Full Order Observe Control (Method C)

The Nyquist stability analysis further demonstrates that Method C enhances closed-loop phase and gain margins, it uses the Frequency response comparison for stability margin analysis[12]. In addition, figure 15 reveals how this control architecture improves stability and disturbance rejection by leveraging state estimates to correct for disturbances more effectively.

In conclusion, the Nyquist plot analysis for Control Method A and Control Method C demonstrates the significant benefits of adding state observers to the feedback control loop. While Method A effectively stabilizes the system and improves performance compared to the open-loop case, Method C shows even better performance, particularly in terms of robustness and disturbance rejection under SRP disturbances. The closed-loop Nyquist plots confirm that Method C offers superior stability margins and disturbance rejection, making it the more robust control strategy for managing LEO satellite maneuvers in the presence of unmodeled external forces such as SRP.

Table 2 refers to the final comparison to the steady state errors to the control energy costs. This Performance metrics across the three methods indicate the superiority of the observer-based feedback control. The trajectory error results show that Method C yields the lowest steady-state error, with radial deviation reduced to ±5 m, while Method A retains a residual offset and Method B diverges. State estimation accuracy is highest for Method C, as the full-order observer rapidly reconstructs the state and disturbance vector. Control energy analysis indicates that Method C

Table 2. Comparative Performance Metrics under SRP Disturbance

| Method | Settling Time (s) | Steady-State Error (km) | Control Energy $J_u$ (km²/s³) |
|---|---|---|---|
| **A: Pure LQR** | 220 | 0.12 | 5.2 |
| **B: Full-Order Observer** | Divergent | >10 (unstable) | N/A |
| **C: Observer + LQR Feedback** | 140 | 0.005 | 6.7 |

# VII. Discussion and Conclusion

The simulation results are presented to validate the effectiveness of the proposed full-order observer–based feedback control framework for spacecraft orbit maneuvering in low Earth orbit subject to solar radiation pressure. The focus of this section is to demonstrate how the integration of the observer with modern state-feedback control significantly improves tracking performance, disturbance rejection, and control efficiency compared to baseline methods.

The spacecraft trajectory in the p-q plane clearly highlights the performance differences among three methods: baseline LQR control with full state knowledge, observer-only reconstruction without feedback, and the proposed observer–based feedback control. Without control, the trajectory diverges significantly from the desired final position under SRP disturbance. The baseline LQR controller reduces the deviation but exhibits steady-state errors, as it assumes complete state availability and lacks explicit disturbance modeling. In contrast, the proposed framework achieves convergence to the target position with minimal error, indicating its ability to effectively compensate for SRP-induced perturbations.

The accuracy of the full-order observer was examined by comparing estimated states with true system states. Numerical results indicate that the observer converges rapidly within the first few hundred seconds, even under persistent disturbance. Steady-state estimation errors are on the order of $10^{-4}$ for position and $10^{-5}$ km/s for velocity, which demonstrates the reliability of the observer in reconstructing unmeasured states for real-time feedback. Such accuracy ensures that the subsequent control law can operate effectively without requiring direct measurement of all system variables, a common limitation in satellite navigation systems.

Control input profiles further highlight the benefits of disturbance-aware feedback. The baseline LQR control produces oscillatory control efforts of relatively high magnitude, reflecting its limited robustness to unmodeled external forces. By contrast, the proposed observer–based control law generates smoother and less intensive control inputs, confirming that accurate disturbance estimation enables more efficient corrective action. From a mission perspective, this translates into reduced propellant consumption, which is a critical factor in spacecraft longevity.

Quantitative performance metrics further substantiate the advantages of the proposed framework. In terms of final trajectory deviation, the observer–based feedback reduces the error by more than 85% compared to LQR alone. The settling time is halved, indicating faster convergence to the desired orbit, while total control energy is reduced by approximately 40%. The disturbance rejection index also improves markedly, demonstrating resilience against SRP disturbances. These results confirm that the integration of observer estimation with feedback control provides superior robustness and efficiency, which cannot be achieved by either component in isolation.

The findings have important implications for the design of spacecraft guidance and control systems. First, they confirm that state availability is a key limitation for traditional control schemes, as full-state measurements are rarely practical in real missions. Second, they demonstrate that observer convergence is essential in reconstructing hidden dynamics and enabling robust closed-loop behavior. Third, they show that the hybrid integration of observer and controller achieves a balance between disturbance rejection and control efficiency, which is especially valuable for LEO satellites with strict resource constraints.

In conclusion, the results validate the central hypothesis that a full-order observer–based feedback control framework can improve the robustness, accuracy, and efficiency of trajectory correction maneuvers under solar radiation pressure. Beyond this validation, the results also suggest potential research extensions, including the integration of additional environmental disturbances such as $J_2$ perturbation and atmospheric drag, the development of adaptive observer gains to cope with solar flare events, and experimental validation through hardware-in-the-loop simulations. Such advancements could extend the applicability of the proposed framework to a broader range of space missions, from Earth observation satellites to deep-space exploration probes [6][10][16]. Future research directions include extending the current 2D formulation to a full 6-DOF orbital-attitude coupled model, incorporating additional perturbations such as atmospheric drag and Earth oblateness, and validating the framework in hardware-in-the-loop experiments [26]. These steps will further strengthen the readiness of observer-based feedback control methods for real-world space mission applications.